\documentclass[12pt,preprint]{aastex}
\usepackage{rotating}
\usepackage{amsmath}

\newcommand{\citar}[1]{\citeauthor{#1} (\citeyear{#1})}
\newcommand{\citarwiths}[1]{\citeauthor{#1}'s (\citeyear{#1})}

\newcommand{\citartwo}[2]{\citeauthor{#1} (\citeyear{#1}, \citeyear{#2})}

\newcommand{\citarNP}[1]{\citeauthor{#1} \citeyear{#1}}

\newcommand{\citartwoNP}[2]{\citeauthor{#1} \citeyear{#1}; \citeyear{#2}}

\begin{document} \title{Observation and modeling of anomalous CN polarization profiles produced by the molecular Paschen-Back
effect in sunspots}

\author{A. Asensio Ramos\altaffilmark{1,2}, J. Trujillo Bueno\altaffilmark{1,3} and M. Collados\altaffilmark{1}}
\altaffiltext{1}{Instituto de Astrof\'{\i}sica de Canarias, 38205, La Laguna, Spain} \altaffiltext{2}{Istituto Nazionale di
Astrofisica (INAF) Osservatorio Astrofisico di Arcetri, Largo Enrico Fermi 5, 50125 Florence, Italy} \altaffiltext{3}{Consejo
Superior de Investigaciones Cient\'{\i}ficas, Spain} \email{aasensio@arcetri.astro.it, jtb@iac.es, mcv@iac.es}

\begin{abstract}
We report novel spectropolarimetric observations of sunspots carried out with the Tenerife Infrared Polarimeter
(TIP) in a near-IR spectral region around 15410 \AA, which is known to contain two groups of prominent OH lines that show circular
polarization signals of opposite polarity. Surrounding these well-known OH lines, we have discovered the presence of CN lines
of the $\Delta v=1$
%$v$=$0-1$, $v$=$1-2$ and $v$=$2-3$
band which show
anomalous polarization profiles. Although the Stokes $V$ signals of the OH lines are antisymmetric and with a sizable amplitude, the
CN lines show almost negligible circular polarization. On the contrary, the linear polarization signals turn out to be much
stronger in the CN lines than in the OH lines. Interestingly, these CN lines present striking antisymmetric linear polarization
profiles, which we are able to explain and model
via the Paschen-Back effect theory for diatomic molecules.
The presence of such peculiar CN lines in the same spectral region of the OH lines may be useful to improve our empirical knowledge
of solar magnetic fields via the simultaneous observation and modeling of the transverse and longitudinal Zeeman effects
in two different molecular species. \end{abstract}


%%%%%%%%%%%%%%%%%%%%%%%%%%%%%%%%%%%%%%%%
%%%%%%%%%%%%%%%%%%%%%%%%%%%%%%%%%%%%%%%%
% INTRODUCTION
%%%%%%%%%%%%%%%%%%%%%%%%%%%%%%%%%%%%%%%%
%%%%%%%%%%%%%%%%%%%%%%%%%%%%%%%%%%%%%%%%
\section{Introduction} The observation and theoretical modeling of the Zeeman effect in molecular lines is a frequently overlooked
diagnostic window, which should however be increasingly pursued because it can lead to important new insights in solar and stellar
physics. The basic theory of the Zeeman effect for doublet states of diatomic molecules was developed a long time ago
(\citarNP{hill29}), but it remained ``forgotten'' for almost half a century, probably due to the difficulty in dealing with its
complicated expressions obtained using basis functions from Hund's case (b)
%for the angular momentum coupling of electronic and
%rotational motion.
In the seventies, \citar{schadee78} developed the
theory of the Zeeman effect using basis functions from Hund's case (a), thus leading to much more simple formulae. However, after an
investigation by \citar{illing81} aimed at explaining the broad-band circular polarization observed by \citar{harvey73} in sunspots
in the $v$=$0-0$ band of the red system ($A^2\Pi-X^2\Sigma$) of CN, the theory of the Zeeman effect in molecular lines was not applied
to solar observations for more than 20 years. Recently, molecular spectropolarimetry has become again an active field of research in
which new observations are being performed and modeled in terms of the Zeeman effect theory
(\citartwoNP{berdyugina00}{berdyugina03}, \citartwoNP{berdyugina01}{berdyugina02a}, \citarNP{asensio_trujillo_spw3_03},
\citarNP{uitenbroek_asensio04}, \citartwoNP{asensio_ch04}{asensio_trujillo04}).

Here we report full Stokes vector observations of sunspots in the same spectral region where one can find the two pairs of OH lines
whose circular polarization profiles were first observed by \citar{harvey85} and modeled by \citar{berdyugina01}. As we shall
see below, \citarwiths{harvey85} OH
lines are surrounded by several CN lines
of the $\Delta v=1$
%$v$=$0-1$, $v$=$1-2$ and $v$=$2-3$
band whose \emph{linear} polarization profiles present an anomalous behavior.

%%%%%%%%%%%%%%%%%%%%%%%%%%%%%%%%%%%%%%%%
%%%%%%%%%%%%%%%%%%%%%%%%%%%%%%%%%%%%%%%%
% OBSERVATIONS
%%%%%%%%%%%%%%%%%%%%%%%%%%%%%%%%%%%%%%%%
%%%%%%%%%%%%%%%%%%%%%%%%%%%%%%%%%%%%%%%%
\section{Observations}
On 30 July 2000 we carried out spectropolarimetric observations of two near-IR spectral regions using the
Tenerife Infrared Polarimeter (TIP, \citarNP{martinez_pillet99}) mounted on the German Vacuum Tower Telescope (VTT) at the
Observatorio del Teide (Iza\~na, Spain). We were interested in accurately measuring the Stokes profiles induced by the
Zeeman effect in the four prominent OH lines that were first observed by Harvey (1985). To our surprise we discovered that, in
addition to such well-known OH lines, there were several enigmatic polarization signals whose identification and radiative transfer
modeling is the main contribution of this Letter.

\begin{figure}
\centering
\plotone{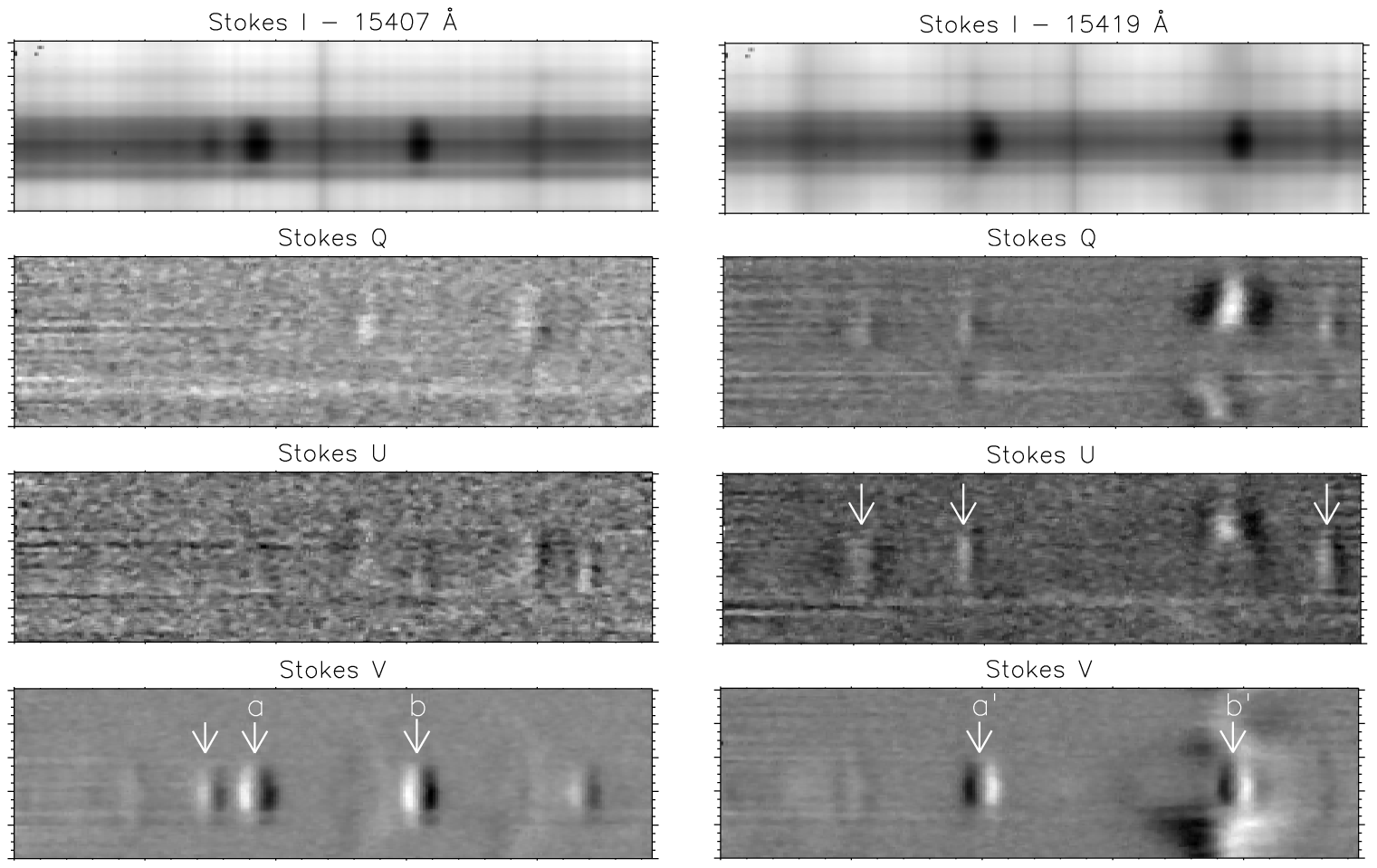}
\caption{Observed Stokes profiles in a sunspot umbra located at $\mu$=$\cos \theta$=$0.75$,
with $\theta$ the heliocentric angle. The lines showing conspicuous circular polarization Stokes $V$ signals belong to OH. Note that
the Stokes $V$ profiles of the two strongest OH lines at lower wavelengths (left panel) have the opposite polarity to those
corresponding to the two strong OH lines at longer wavelengths (right panel). The lines showing weak antisymmetric linear
polarization belong to CN. Wavelength increases to the right and the vertical direction is the spatial direction along the slit. The
OH lines are marked with arrows in the Stokes $V$ image while the CN lines are marked with arrows in the Stokes $U$ image. The
$(a,b)$ and $(a',b')$ pairs of lines are \citarwiths{harvey85} OH lines. The signal to the left of the $(a,b)$ pair in Stokes $V$
belongs also to OH.
%The one to the right remains unidentified, although we think it might be associated to CN.
}
\label{fig_cn_oh_observation}
\end{figure}

Figure \ref{fig_cn_oh_observation} shows the recorded spectra in the two spectral regions we observed around 1.54 $\mu$m. The slit
was crossing a sunspot situated at $\mu$=$\cos \theta $=$0.75$, with $\theta$ the heliocentric angle. The four OH lines observed by
\citar{harvey85} are vibro-rotational lines of the $v$=$2-0$ system arising in the fundamental electronic state $X^2\Pi$. They
can be seen in the figure as strong absorptions in the
sunspot umbra, which completely disappear outside the sunspot. Even in the penumbral spectrum the absorption in the OH lines is
drastically reduced. Concerning the circular polarization spectrum, we have verified that the OH lines present conspicuous
antisymmetric Stokes $V$ profiles. Such two pairs of OH lines are the $(a,b)$ and $(a',b')$ circular polarization features of Fig.
\ref{fig_cn_oh_observation}, whose Stokes V profiles have opposite
sign. As shown in Fig. \ref{fig_cn_oh_observation}, in addition
we have detected two extra circular polarization signals situated to the left and to the right of Harvey's (1985) $(a,b)$ pair of OH
lines. The left one is the P$_2$(13/2) transition of OH, while we are presently working on the exact identification
of the spectral feature located to the right hand side.

Additionally, some lines showing very conspicuous linear polarization profiles (even at the center of the sunspot umbra) are also present
in this spectral region although they show very weak absorptions in the intensity
spectrum. They show also very weak Stokes $V$ signals. Note that although the
above-mentioned OH lines have a relatively high value of the effective Land\'e factor, they do not show clean and sizable
Stokes $Q$ and/or $U$ signals.

A detailed comparison of the exact wavelengths of these lines with the CN linelist of \citar{kurucz93} indicates that they
are CN lines. We find that the CN linear polarization features are more conspicuous close to the
umbra-penumbra border, while the observed amplitudes become very small at the center of the umbra. Furthermore, the region of the
sunspot which is closer to the disk center shows weaker polarization signals produced by these CN transitions. Additionally, the
fact that the linear polarization in such CN lines is much more significant than the circular polarization is common to all the
points in the sunspot.

Even more striking than the previous considerations is that the linear polarization of our CN lines do not present the
typical symmetric profiles of the transverse Zeeman effect. In fact, the shape closely resembles that of the typical antisymmetric
profiles of the circular polarization observed in the OH lines.
As we shall see below, this curious behavior is due to the
molecular Paschen-Back effect.
To the best of our knowledge, this is the first time that
antisymmetric {\em linear} polarization profiles induced
by the Paschen-Back effect are clearly observed in molecular
lines\footnote{It is of interest to mention that in his theoretical
paper \citar{illing81} showed a figure with
computed Stokes $V$ and $Q$ profiles for the $v=0-0$
$R_2(13)+{^{R}}Q_{12}(13)$ complex where
one can see more amplitude for $V$ than for $Q$. This behavior is clearly different to what we have
found both observationally and theoretically for the $v$=$0-1$, $v$=$1-2$ and $v$=$2-3$ bands.}.

%%%%%%%%%%%%%%%%%%%%%%%%%%%%%%%%%%%%%%%%
%%%%%%%%%%%%%%%%%%%%%%%%%%%%%%%%%%%%%%%%
% PHYSICAL INTERPRETATION
%%%%%%%%%%%%%%%%%%%%%%%%%%%%%%%%%%%%%%%%
%%%%%%%%%%%%%%%%%%%%%%%%%%%%%%%%%%%%%%%%
\section{Physical Interpretation}

%%%%%%%%%%%%%%%%%%%%%%%%%%%%%%%%%%%%%%%%
% OH LINES
%%%%%%%%%%%%%%%%%%%%%%%%%%%%%%%%%%%%%%%%
%\subsection{OH lines}
The reason why the circular polarization of the two pairs of Harvey's (1985) OH lines present opposite
polarity is because they have equal but opposite effective Land\'e factors (\citarNP{ruedi95}, \citarNP{berdyugina01}).
These OH lines are
vibro-rotational lines of the $v$=$2-0$ system arising in the fundamental electronic state $X^2\Pi$. The $\Lambda$-doublet
P$_{1e}(21/2)$ and P$_{1f}(21/2)$ is located around 15419 \AA\ and the $\Lambda$-doublet P$_{2e}(19/2)$ and P$_{2f}(19/2)$ lies
around 15407 \AA\ . Such OH lines are marked in Fig. \ref{fig_cn_oh_observation} as $a$, $b$, $a'$ and $b'$, respectively. Although
the $X^2\Pi$ state has to be described under intermediate coupling, we can safely describe it using Hund's case (b) coupling
because the total angular momentum $J$ of both lines is high enough (cf. \citarNP{berdyugina02a}). Using the formula for the Land\'e factor
in Hund's case (b) (\citarNP{landau_lifshitz82}) and the well-known formula for the effective Land\'e factor (e.g.,
\citarNP{landi92}), we end up with a change of sign for transitions between levels with $J$=$N+1/2$ ($N$ is the total angular momentum
apart from spin) and those between levels with $J$=$N-1/2$:
\begin{equation}
\bar g = \pm \frac{4J^2+8J+1}{2(J+1)(2J+1)(2J+3)},
\end{equation}
where the plus sign corresponds to transitions
between levels $J$=$N+1/2$ and the minus sign to transitions between levels $J$=$N-1/2$. The other (weaker) OH line to the left of the
OH line labeled by ``$a$'' in Fig. \ref{fig_cn_oh_observation} confirms this result since it is the P$_2$(13/2) line.

%%%%%%%%%%%%%%%%%%%%%%%%%%%%%%%%%%%%%%%%
% CN LINES
%%%%%%%%%%%%%%%%%%%%%%%%%%%%%%%%%%%%%%%%
%\subsection{CN lines} \label{sec_cn_lines}
The CN lines we have observed around 1.54 $\mu$m belong to the $A^2\Pi-X^2\Sigma^+$
electronic transition. The rotational levels of the lower electronic state can be correctly described under Hund's case (b) coupling
when we are in the Zeeman regime. The transition to the Paschen-Back regime occurs for very weak fields ($\sim$77 G for the lowest
$J$ rotational levels). For this reason, these lines are always in the Paschen-Back regime under the typical magnetic fields of
sunspots. The rotational levels of the upper electronic state are always in the Zeeman regime because the field
strength at which the transition to Paschen-Back occurs is $\sim$560 kG. However, due to the spin uncoupling, the rotational
levels of the $A^2\Pi$ electronic state
has to be described in an
intermediate coupling between Hund's case (a) and (b).
The theory of the molecular Zeeman effect developed by \citar{schadee78} (since the lines belong to a transition between doublet states)
or the more general numerical diagonalization approach of \citartwo{asensio_trujillo_spw3_03}{asensio_trujillo05b} have to be used
for determining the magnetic properties of the CN lines around 1.54 $\mu$m.

\begin{figure}
\centering
\plottwo{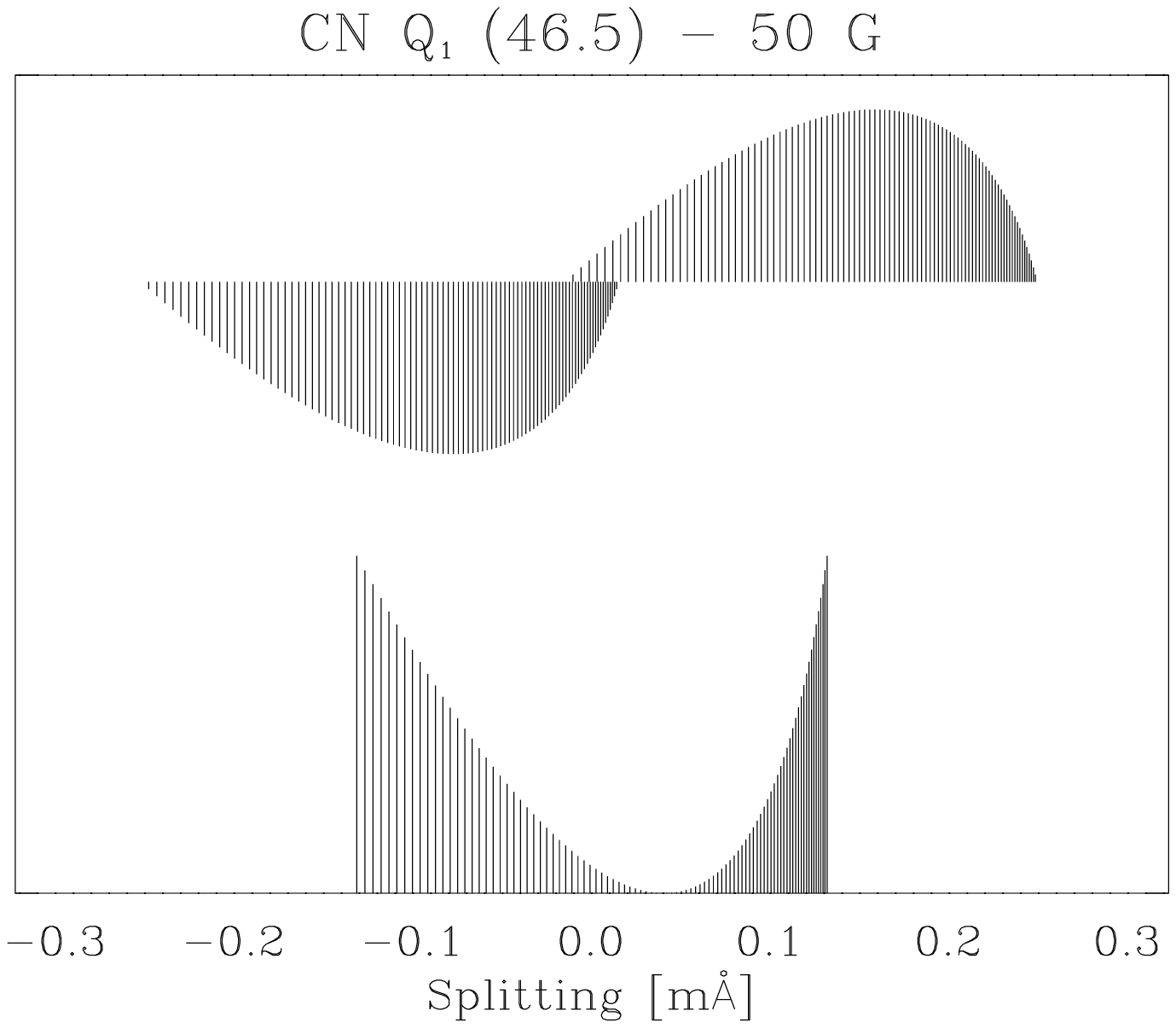}{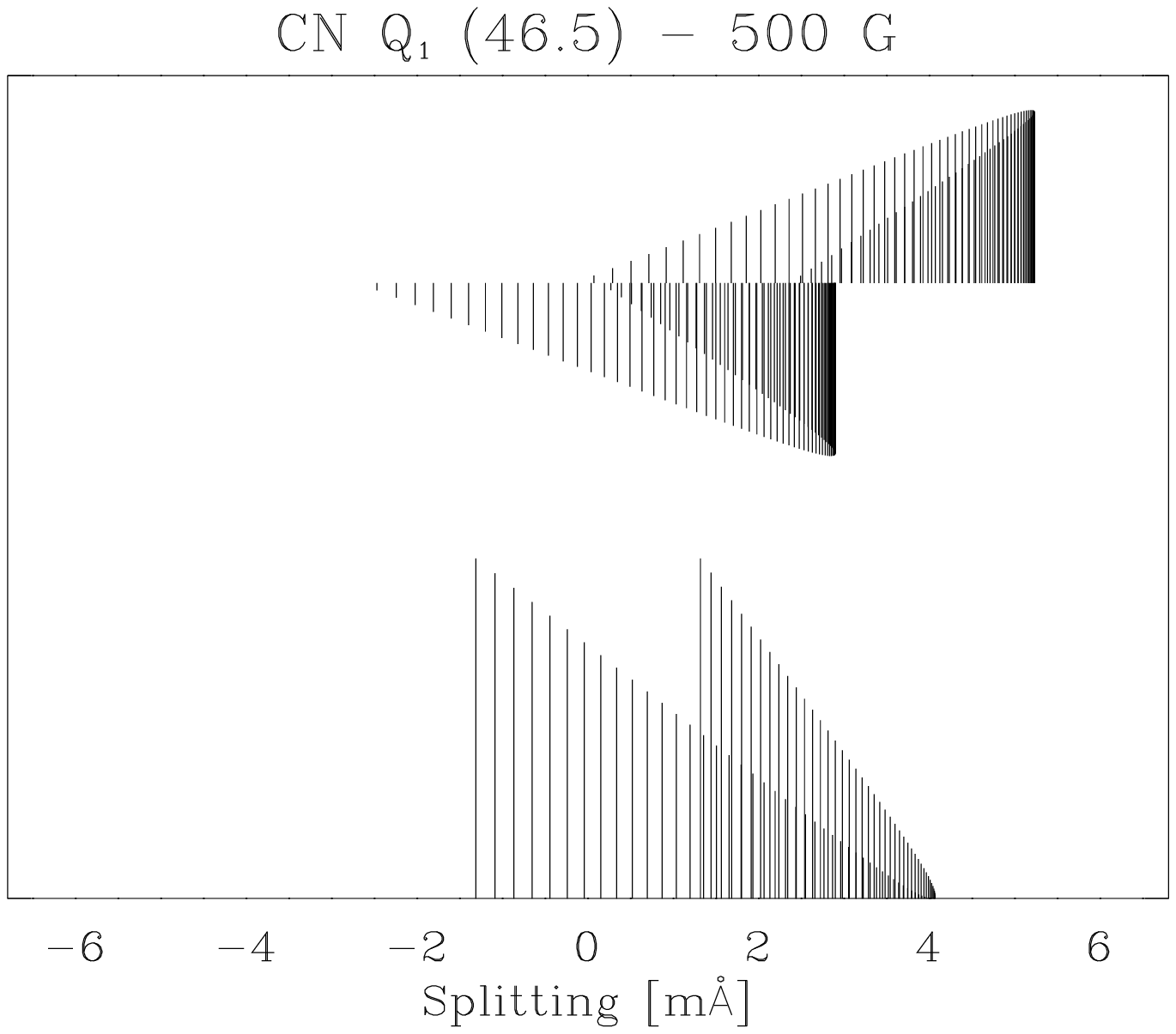}
\plottwo{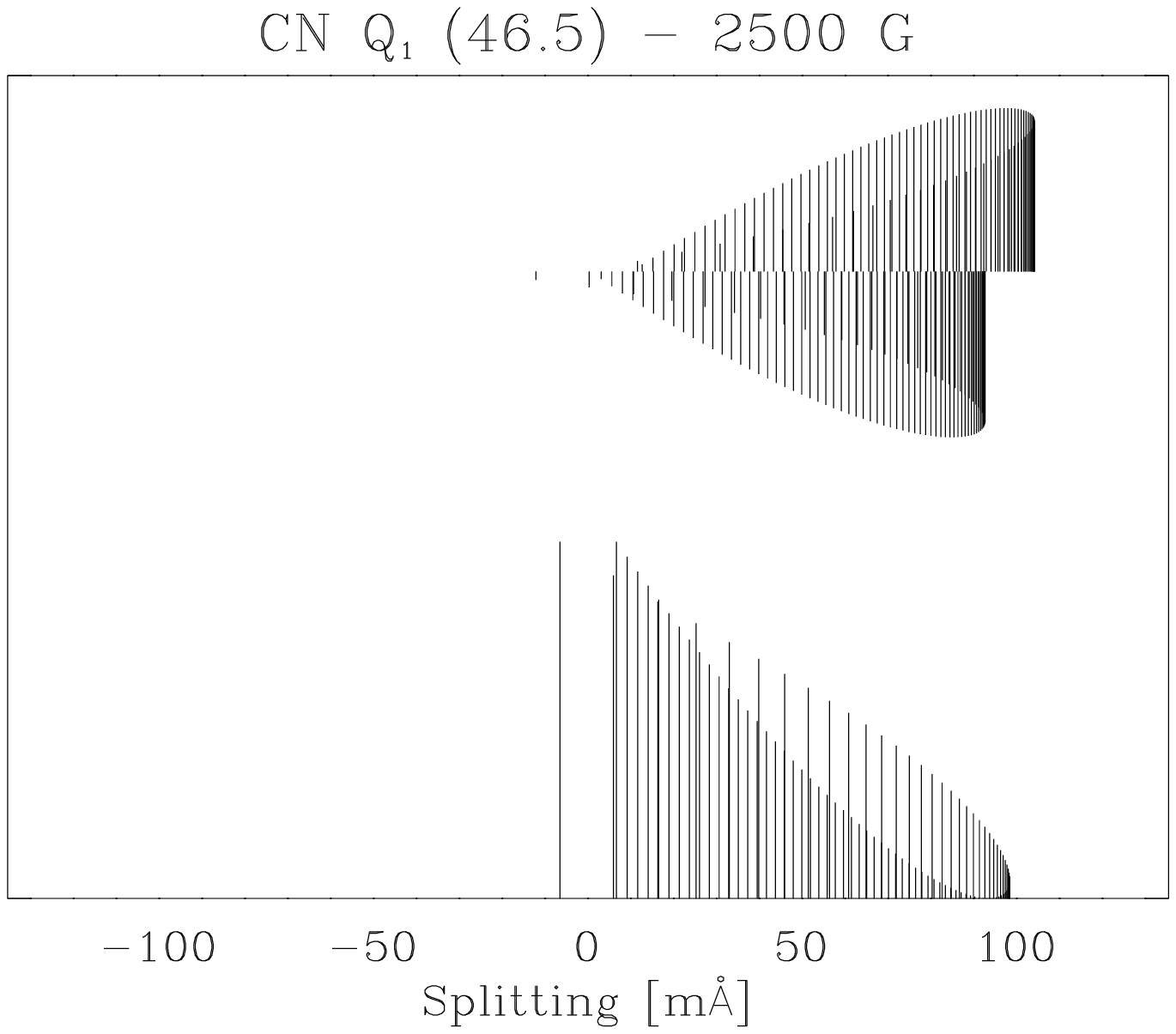}{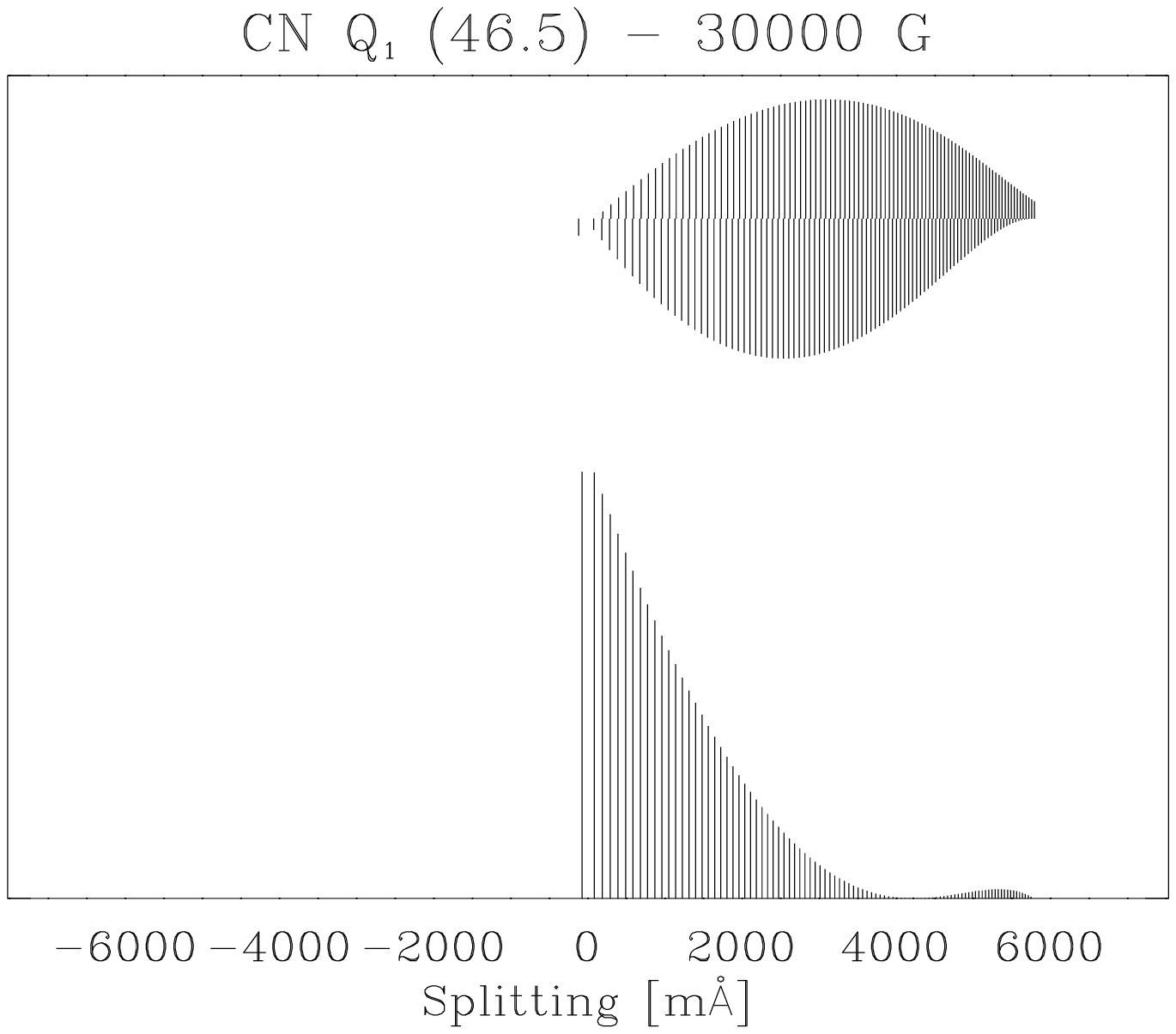}
\caption{Zeeman patterns for 50, 500, 2500 and 30000
G of the Q$_1$(46.5) CN line. Note that the $\sigma$ and $\pi$ components are symmetric for low magnetic field strengths and that
the Zeeman pattern changes drastically as the field strength increases due to the transition to the Paschen-Back regime. At very
high fields, the two $\sigma$ components behave similarly.} \label{fig_cn_patterns}
\end{figure}

The exact wavelength of the observed CN lines are 15418.29 \AA,
15419.27 \AA\ and 15423.39 \AA. These lines arise from transitions between levels
with high values of the angular momentum $J$. The first one is the Q$_1$(27.5) line of the $v$=$2-3$ band ($J_u$=$J_l$=$55/2$ and
$N_u$=$N_l$=$27$), the second is the Q$_1$(59.5) of the $v$=$0-1$ band ($J_u$=$J_l$=$119/2$ and $N_u$=$N_l$=$59$) and the third one is the
Q$_1$(46.5) line of the $v$=$1-2$ band ($J_u$=$J_l$=$93/2$ and $N_u$=$N_l$=$46$). We show in Fig. \ref{fig_cn_patterns} the Zeeman
patterns for the Q$_1$(46.5) line for
different field strengths. These Zeeman patterns indicate the wavelength shift and strength of each of the Zeeman components of the
transition. The upper part shows the two $\sigma$ components: those arising from $\Delta M$=$-1$ transitions going upwards and those
arising from $\Delta M$=$1$ going downwards. The lower part shows the $\pi$ component, i.e., transitions having $\Delta M$=$0$.
The rotational and coupling constants have been obtained from \citar{huber_herzberg03}.

%The rotational and coupling constants, which have been obtained from \citar{huber_herzberg03}, are
%$B_\mathrm{rot}(X^2\Sigma^+)$=$1.8997$ cm$^{-1}$,
%$B_\mathrm{rot}(A^2\Pi)$=$1.7151$ cm$^{-1}$, $\gamma_\mathrm{S-R}(X^2\Sigma^+)$=$7.25\times 10^{-3}$ cm$^{-1}$ and
%$A_\mathrm{SO}(A^2\Pi)$=$-52.64$ cm$^{-1}$.

Figure \ref{fig_cn_patterns} shows that, even for fields as low as 50 G, the Zeeman patterns of these CN transitions are slightly
perturbed by the Paschen-Back effect. This transition to the Paschen-Back regime at low fields is produced because the Zeeman
splitting for such low fields is of the order of the energy separation between two consecutive levels of the lower electronic state.
Therefore, the non-diagonal terms in the Hamiltonian between adjacent $J$ levels are non-negligible. When the field strength is
increased, the perturbation becomes larger until arriving to field strengths as large as 30000 G, where the two $\sigma$ components
are converging to the same structure.
%What we have just described is nothing but the transition from the Zeeman to the incomplete
%Paschen-Back regime.

Although it is necessary to solve the radiative transfer equation for polarized light in order to obtain the emergent Stokes
profiles, it suffices to consider the Zeeman patterns for the case of 2500 G for understanding their qualitative shape.
%Since the observations were made on
%a sunspot, we focus now on the Zeeman pattern obtained for a magnetic field of 2500 G.
The final profile is obtained by adding
%In order to obtain the final line profile, one has to add
many Voigt profiles centered at each vertical line in the Zeeman pattern
diagram of Fig. \ref{fig_cn_patterns}
with a relative strength given by the length of the vertical line. However, we can obtain a general idea of the shape of the
$\phi_q$ profiles if we consider a Voigt profile centered at the center of gravity of each component. This way, $\eta_V$ is
proportional to the difference between two profiles centered at the two $\sigma$ components, while $\eta_Q$ is proportional to the
difference between a profile centered in the $\pi$ component and the average of two profiles centered in the $\sigma$ components.
Note in Fig. \ref{fig_cn_profiles} that the centers of gravity of the two $\sigma$ components tend to converge to the same
wavelength shift when the magnetic field strength is increased, therefore leading to a smaller value of $\eta_V$. This represents
the fundamental reason for the observations not showing strong Stokes $V$ profiles in the CN lines even though the magnetic field is
mainly vertical in the observed sunspot umbra. Using the same reasoning, it is immediately explained why Stokes $U$ is stronger than
Stokes $V$, and why Stokes $U$ presents an antisymmetric profile. To the best of our knowledge, this antisymmetric Stokes $U$ profile
produced by the underlying structure of the Zeeman pattern itself has not been observed in atomic lines.

%When the magnetic field vector lies along the line-of-sight, the polarization signal in these CN lines is only circular but very
%weak. On the contrary, when the magnetic field vector is inclined with respect to the line-of-sight, the linear polarization signal
%rapidly increases, becoming larger than the circular polarization for barely inclined fields.

%%%%%%%%%%%%%%%%%%%%%%%%%%%%%%%%%%%%%%%%
% STOKES PROFILES
%%%%%%%%%%%%%%%%%%%%%%%%%%%%%%%%%%%%%%%%
\section{Theoretical Stokes Profiles} 
\begin{figure}
\centering
\plottwo{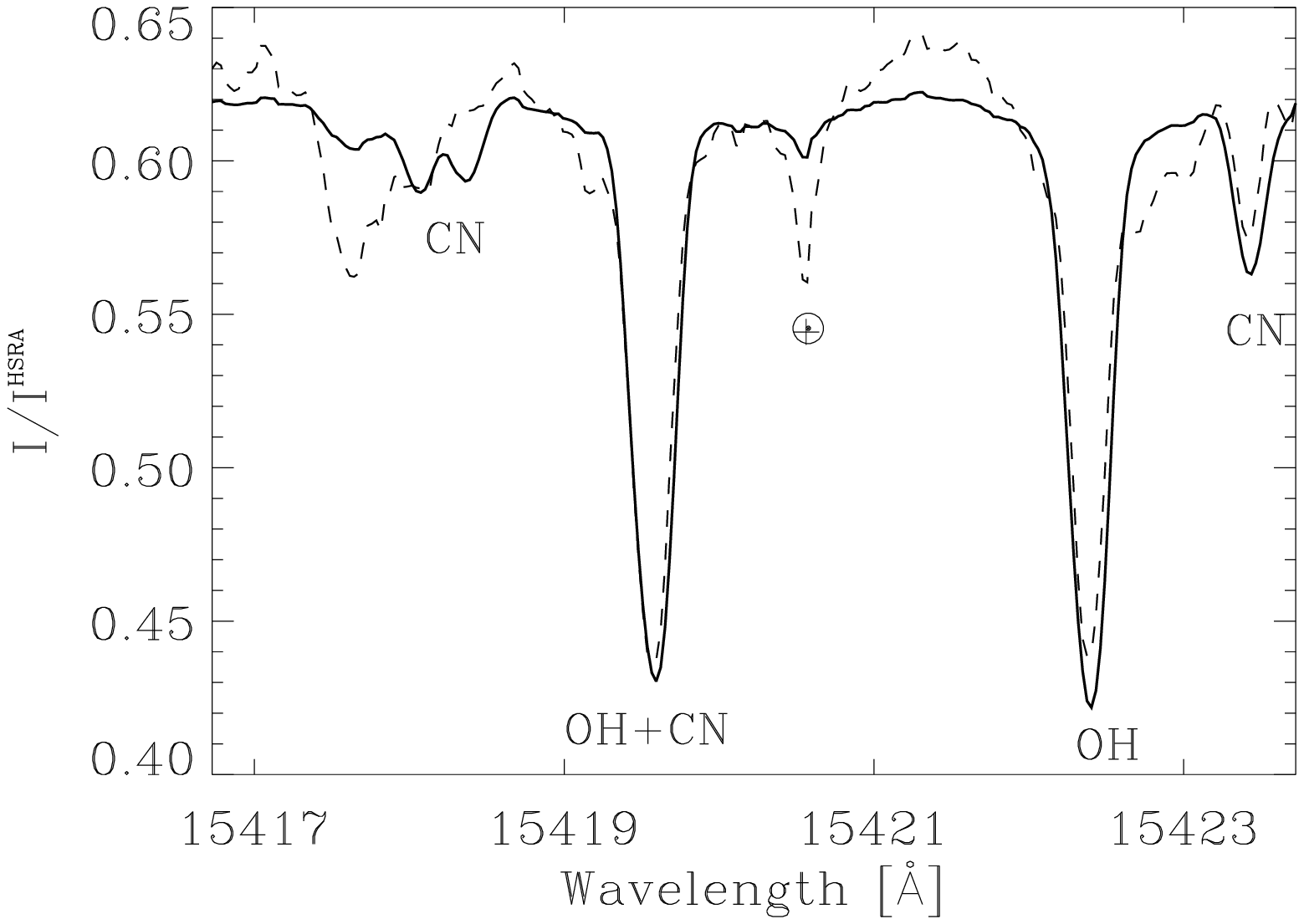}{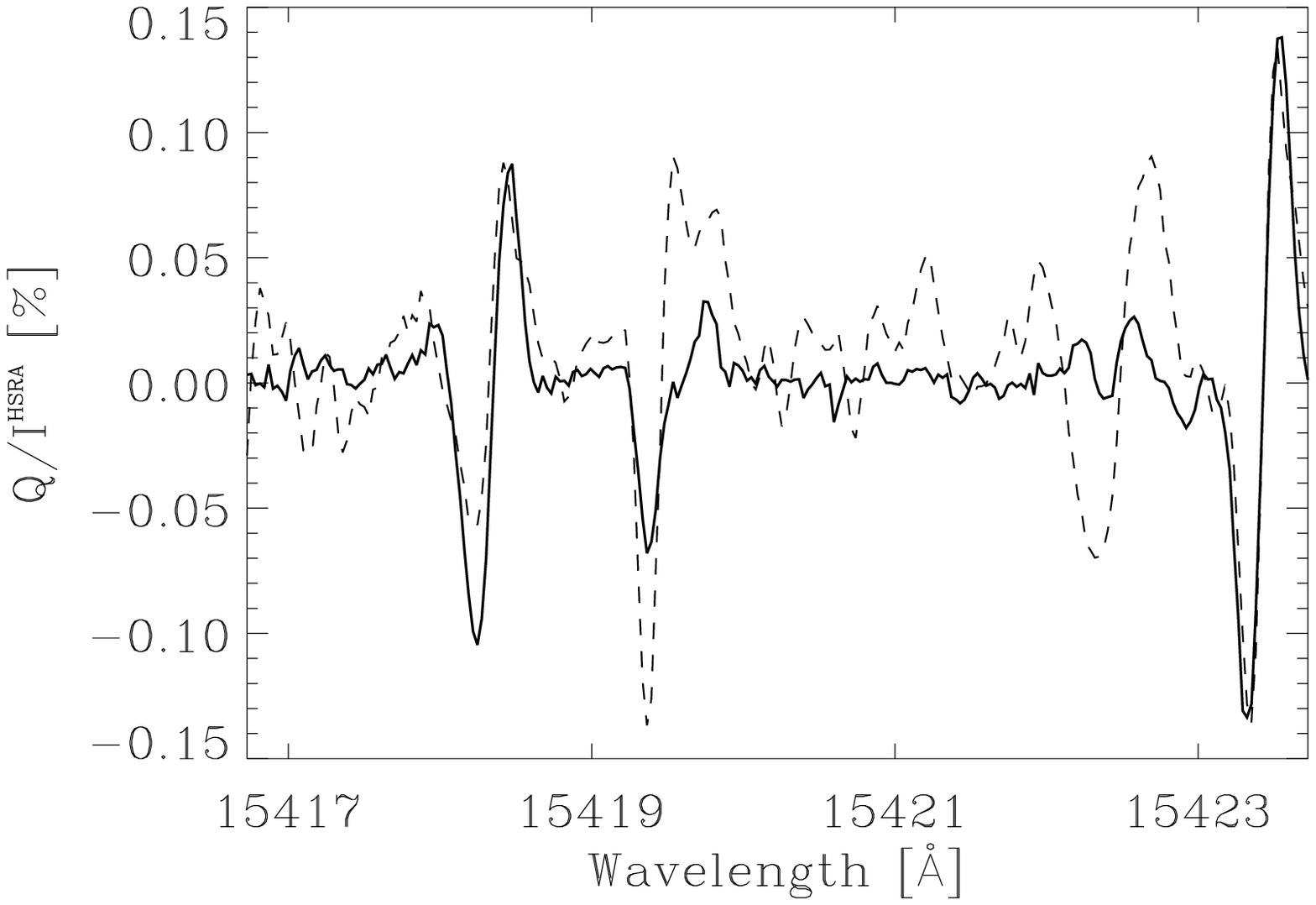}
\plottwo{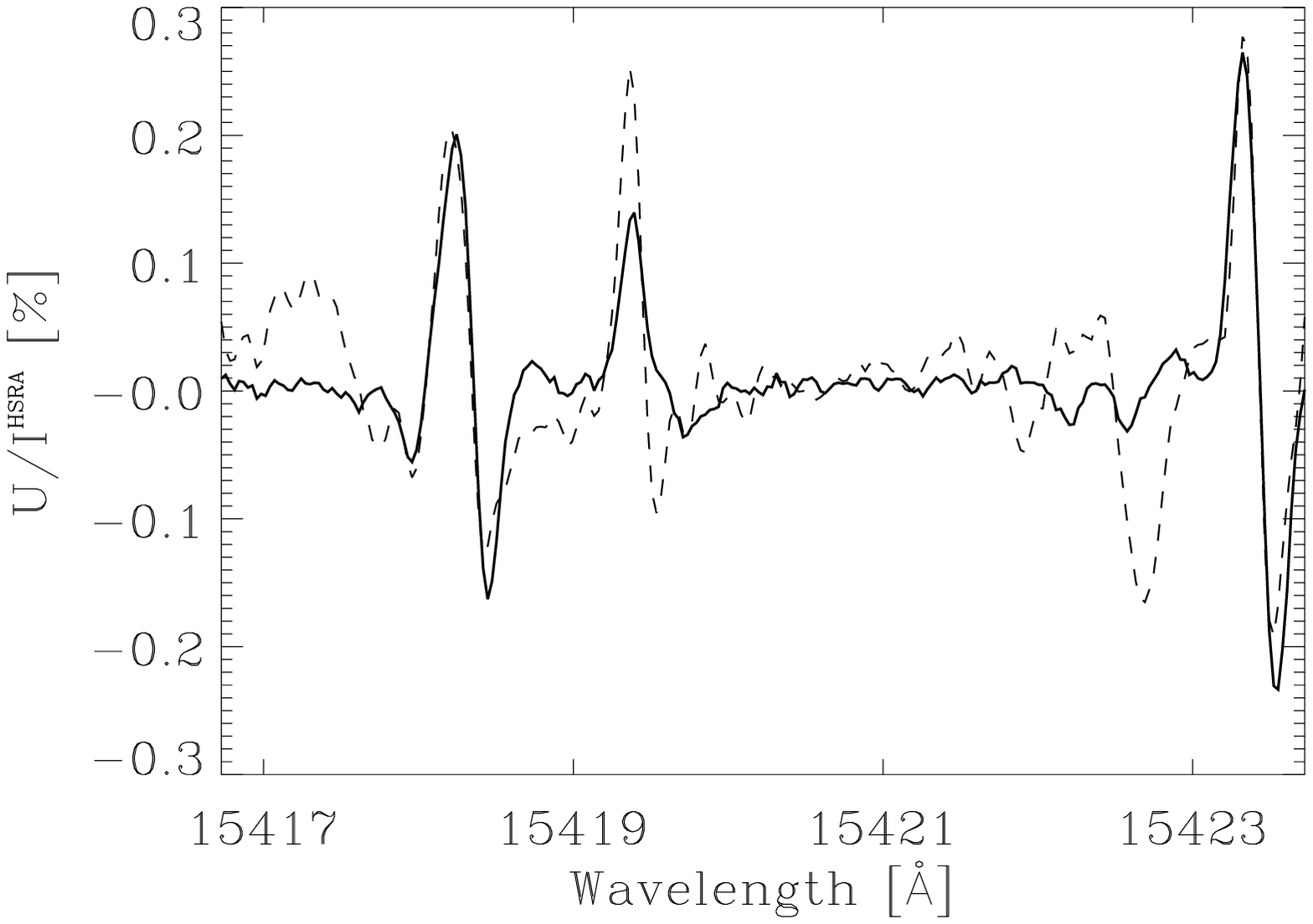}{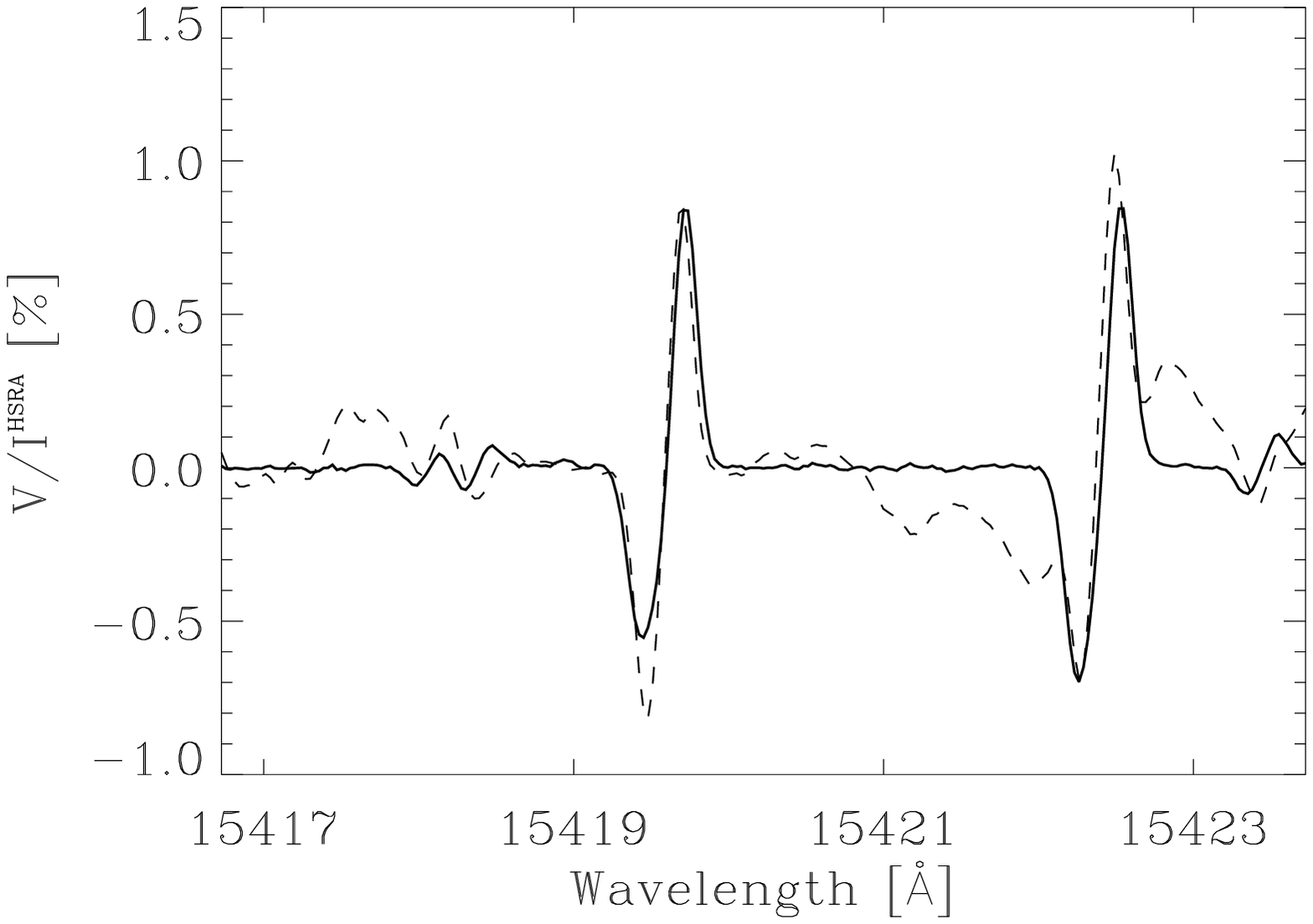} \caption{Stokes profiles observed in the center of
the umbra (dotted line) and the synthetic profiles obtained by solving the radiative transfer equation in a model atmosphere
obtained after an inversion of the observed Stokes profiles (solid lines).
Both, the observed and synthetic profiles are normalized to the continuum intensity
calculated in the Harvard-Smithsonian Reference Atmosphere (HSRA, \citarNP{gingerich71}).
Note that the shapes of the linear polarization profiles of the CN lines
are fairly well fitted.} \label{fig_cn_profiles}
\end{figure}

In order to show that the previous qualitative discussion based on Zeeman patterns
arguments is indeed correct, we have synthesized the whole spectral region using our radiative transfer code for polarized
radiation, which solves the Stokes vector transfer equation
via the quasi-parabolic short-characteristics method (\citarNP{trujillo03}).
The model atmosphere we have used
has been obtained from the application of an inversion code of
Stokes profiles induced by the molecular Zeeman effect, which will be described elsewhere. Figure \ref{fig_cn_profiles} shows the
observed Stokes profiles normalized to the surrounding quiet Sun spectrum (dotted line) and the synthetic profiles. Concerning Stokes $I$, we recover with fairly good accuracy the depth and width of the strong OH lines, while the CN
lines present a worse fit. Concerning the Stokes $V$
profiles of the OH lines, we recover the typical antisymmetric shape, but with a slight asymmetry produced by the individual Zeeman
patterns. The Stokes $V$ signals produced by the CN lines are very weak and we cannot make a direct comparison between observation
and modeling.

The
%linear polarization signals
Stokes $Q$ and $U$
profiles clearly show the
antisymmetric signals produced by the CN lines.
The feature at 15419.27 \AA\ is produced by the blend between OH and CN, which is difficult to fit. In fact, this
blend is a massive one in which many CN lines are at similar wavelengths than the OH line and the misfit might be produced by
incorrect values of the line strengths. This is reinforced by the fact that the general shape of the profile appears to be recovered
in the synthetic profile. In order to avoid these uncertainties,
we focus on the cleaner CN lines.
%it is better to focus on the clean CN lines which show conspicuous
%linear polarization signals.
Although clean, they indeed represent the contribution of several weak CN lines surrounding the strong
component which produces the main signal.

The Stokes $Q$ profile of the CN line at 15423.39 \AA\ shown in Fig. \ref{fig_cn_profiles} is very well reproduced. This line
presents two lobes of about the same strength, which is correctly recovered using the corresponding Zeeman patterns.
Stokes $U$
%shows again an almost completely antisymmetric profile with two lobes of the same strength.
presents a very similar structure.
This regularity is produced because
%of
%the high value of the $J$ quantum number of the levels of the transition. The
deviations from the Zeeman regime are less important
for lines between levels with high value of $J$ than for levels with low value of $J$ (the spin-rotation splitting in the $^2\Sigma$
state increases when $J$ increases). Concerning the line at 15418.29 \AA\, we find a good fit to Stokes $Q$ and $U$ even although
the line shows quite asymmetric profiles.

%%%%%%%%%%%%%%%%%%%%%%%%%%%%%%%%%%%%%%%%
%%%%%%%%%%%%%%%%%%%%%%%%%%%%%%%%%%%%%%%%
% CONCLUSION
%%%%%%%%%%%%%%%%%%%%%%%%%%%%%%%%%%%%%%%%
%%%%%%%%%%%%%%%%%%%%%%%%%%%%%%%%%%%%%%%%
\section{Conclusions}
We have presented spectropolarimetric observations of a sunspot at $\mu$=$0.75$ taken with the Tenerife Infrared
Polarimeter in a near-IR spectral region which contains the two groups of OH lines that were first observed by Harvey (1985). Such
groups of OH lines show conspicuous {\em circular} polarization signals.
Interestingly, our full Stokes-vector observations reveal the presence of hitherto undetected CN lines
%at 15418.29 \AA, 15419.27 \AA\
%and 15423.39 \AA\,
which show anomalous {\em linear} polarization profiles. The observed Stokes $Q$ and $U$ profiles turn out to be much
stronger than Stokes $V$, even in the presence of the mainly vertical umbral fields that produce the above-mentioned OH circular
polarization. Moreover, the CN linear polarization profiles present antisymmetric shapes which resemble those corresponding to the
typical circular polarization profiles of the Zeeman regime. We have shown that the theory of the molecular Zeeman effect is able to
explain the observed anomalous linear polarization profiles because such CN lines are in the Paschen-Back regime for the typical kG
fields of sunspots.

As mentioned in Section 1, Harvey (1973) had discovered previously the existence of non-zero net {\em circular} polarization
for some lines of the $v$=$0-0$ band of the red system of CN, which were later explained by Illing (1981) in terms of the molecular
Paschen-Back effect. However, to the best of our knowledge, this is the first time that CN lines are identified which, in spite of showing
minor absorption features and negligible circular polarization
amplitudes, show however conspicuous linear polarization profiles whose shape is antisymmetric. Such CN lines are of great diagnostic interest because they happen to be located in the very same spectral
region where the above-mentioned OH lines show sizable circular polarization signals. Among other diagnostic applications in solar
and stellar physics, this offers the possibility of inferring the
three-dimensional structure of the magnetic field vector in sunspots via the simultaneous observation and modeling of the transverse and
longitudinal Zeeman effects in two different molecular species.

\acknowledgments This research has been partly funded by the Ministerio de Educaci\'on y Ciencia through project AYA2004-05792 and by
the European Solar Magnetism Network (contract HPRN-CT-2002-00313).

\end{document}